\documentclass[a4paper,11pt]{article}
\usepackage{jheppub}
\usepackage{mathrsfs}
\usepackage{verbatim}
\usepackage{booktabs}
\usepackage{placeins}
\usepackage{pdflscape}
\usepackage{slashed}

\newcommand{\R}{\mathbb{R}}

\newcommand{\C}{\mathbb{C}}

\newcommand{\nn}{\nonumber}
\newcommand{\mc}[1]{\mathcal{#1}}

\newcommand{\ISP}{\mathbb{ISP}}
\newcommand{\pd}[2]{\frac{\partial #1}{\partial #2}}
\newcommand{\tree}{\mathrm{tree}}
\newcommand{\spaa}[1]{\langle#1\rangle}
\newcommand{\spbb}[1]{[#1]}
\newcommand{\spab}[3]{\langle#1|#2|#3]}

\newcommand{\expval}[3]{\langle #1|\hspace*{.2mm}#2\hspace*{.3mm}|#3 \rangle}
\newcommand{\Res}{\mathop{\rm Res}}
\newcommand{\bc}{\begin{center}}
\newcommand{\ec}{\end{center}}

\newcommand{\X}{X_{1,1,2}^{**}}
\renewcommand{\P}{P_{2,2}^{**}}

\newtheorem{thm}{Theorem}

\title{Unitarity Cuts of Integrals with Doubled Propagators}
\author[a,b]{Mads S{\o}gaard}
\author[a]{Yang Zhang}
\affiliation[a]{
Niels Bohr International Academy and Discovery Center, Niels Bohr Institute, \\
University of Copenhagen, Blegdamsvej 17, DK-2100 Copenhagen, Denmark
}
\affiliation[b]{
Institut de Physique Th\'eorique, CEA-Saclay, F--91191 Gif-sur-Yvette cedex,
France
}
\emailAdd{madss@nbi.dk}
\emailAdd{zhang@nbi.dk}

\abstract{We extend the notion of generalized unitarity cuts to accommodate
loop integrals with higher powers of propagators. Such integrals frequently
arise in for example integration-by-parts identities, Schwinger
parametrizations and Mellin-Barnes representations. The method is applied to
reduction of integrals with doubled and tripled propagators and direct
extract of integral coefficients at one and two loops. Our algorithm is based
on degenerate multivariate residues and computational algebraic geometry.}

\begin{document}
\maketitle
\flushbottom
\clearpage

\section{Introduction}
Perturbative scattering amplitudes of elementary particles in quantum field
theories such as Quantum Chromodyanmics (QCD) are traditionally calculated by
means of Feynman diagrams and rules. The Feynman approach is very intuitive,
but suffers from a severe proliferation of terms and diagrams for increasing
number of external particles and order in perturbation theory. Moreover, the
simplicity of the underlying theory is by no means reflected by the results.
Recent years have seen enormous progress in quantitative determination of
amplitudes at the one-loop level and beyond, catalyzed by the demand of
precise theoretical predictions by the Large Hadron Collider (LHC) programme
at CERN. 
 
Modern computations of scattering amplitudes take advantage of the properties
of analyticity and unitarity. Analyticity allow for amplitudes to be
reconstructed directly from their analytic structure, while unitarity implies
that residues at their singularities factorize onto products of simpler
amplitudes. Advances along these lines include among others the unitarity
method \cite{Bern:1994cg,Bern:1994zx} by Bern, Dixon, Dunbar and Kosower (see
e.g.
refs.~\cite{Bern:1995db,Bern:1997sc,Britto:2004nc,Britto:2004nj,Bern:2005hh,
Bidder:2005ri,Britto:2005ha,Britto:2006sj,Mastrolia:2006ki,Brandhuber:2005jw,
Ossola:2006us,Anastasiou:2006gt,Forde:2007mi,Badger:2008cm,Giele:2008ve,
Britto:2006fc,Britto:2007tt,Bern:2010qa,Anastasiou:2006jv,Bern:1997nh,
Bern:2000dn,Bern:2002tk}) and the Britto-Cachazo-Feng-Witten recursion
relations \cite{Britto:2004ap,Britto:2005fq} for tree amplitudes. In the last
couple of years, new developments in automation of two and three loop
amplitudes calculations in arbitrary gauge theories at the level of an
integral basis
\cite{Kosower:2011ty,Larsen:2012sx,CaronHuot:2012ab,Johansson:2012zv,
Sogaard:2013yga,Sogaard:2013fpa,Johansson:2013sda,Gluza:2010ws}
and also at the integrand
\cite{Badger:2012dp,Mastrolia:2011pr,Badger:2012dv,Zhang:2012ce,
Badger:2013gxa,Feng:2012bm,Mastrolia:2012an,Mastrolia:2012wf,Huang:2013kh,
Mastrolia:2013kca}
have been reported. These papers generalize the works at one-loop of
Britto, Cachazo, Feng \cite{Britto:2004nc}, Forde \cite{Forde:2007mi} and
Ossola, Papadopoulos and Pittau \cite{Ossola:2006us}. We also refer to 
\cite{Glover:2001af,Anastasiou:2000kg,Anastasiou:2000ue,Anastasiou:2001sv,
Buchbinder:2005wp,Cachazo:2008vp,Schabinger:2011dz,Feng:2014nwa}
for more related literature.

In this paper, we define generalized unitarity cuts of integrals that
otherwise appear incompatible with the usual cut prescription. Naively,
applying a unitarity cut to an integral with higher powers of propagators, the
immediate result is singular. Nevertheless, amplitude representations that
contain integrals with doubled propagators can lead to significant
simplifications as argued in \cite{Henn:2013pwa}. Moreover, integrals with
repeated propagators naturally appear in many actual calculations. We explain
that such cuts can be treated as degenerate multivariate residues using
computational algebraic geometry. Our algorithm makes it possible to use more
general integral bases for loop amplitudes. In particular, we provide examples
of two-loop integral bases, whose elements contain purely scalars and yet are
adaptable for unitarity purposes. What is more, the algorithm can be used to
derive integration-by-parts (IBP) identities for integrals with repeated
propagators.

The integrand reduction of two-loop diagrams with doubled propagators has been
achieved in \cite{Mastrolia:2013kca}, via the synthetic polynomial division
method. However, the full integral reduction for integrals with doubled
propagators has not been considered from the unitarity viewpoint.

\subsection{Generalized Feynman Integrals}
We define the generalized dimensionally regularized $n$-loop Feynman integral
with arbitrary integer powers (also called indices)
$(\sigma_1,\dots,\sigma_p)$ of $p$ propagators by
\begin{align}
I(\sigma_1,\dots,\sigma_p) = 
\int\frac{d^D\ell_1}{(2\pi)^D}\cdots
\int\frac{d^D\ell_n}{(2\pi)^D}
\prod_{k=1}^{p}\frac{1}{f^{\sigma_k}_k(\{\ell_i\})}\;,
\end{align}
where the $f_k$'s are linear polynomials with respect to inner products of
the $n$ loop momenta $\{\ell_i\}$ and $m$ external momenta $\{k_i\}$. The
canonical integral is recovered when all indices are set to unity. Generally
speaking, the integral will have a nontrivial numerator function and is in
that case referred to as a tensor integral. We can always bring the numerator
into the form of additional propagators raised to negative powers.

In a typical multiloop amplitude calculation, a huge number of Feynman
integrals with different indices appear. A subset of the integrals can easily
be reduced algebraically, e.g. by means of Gram matrix determinants. At first
glance, the remaining integrals may seem irreducible and independent, but they
are in fact related by IBP identities \cite{Chetyrkin:1981qh}. Discarding the
boundary term in $D$-dimensional integration, total derivatives vanish upon
integration,
\begin{align}
\int\frac{d^D\ell_1}{(2\pi)^D}\cdots
\int\frac{d^D\ell_n}{(2\pi)^D}
\pd{}{\ell_a^\mu}\bigg(
k_b^\mu\prod_{k=1}^{p}\frac{1}{f^{\sigma_k}_k(\{\ell_i\})}
\bigg) = 0\;, \\
\int\frac{d^D\ell_1}{(2\pi)^D}\cdots
\int\frac{d^D\ell_n}{(2\pi)^D}
\pd{}{\ell_a^\mu}\bigg(
\ell_b^\mu\prod_{k=1}^{p}\frac{1}{f^{\sigma_k}_k(\{\ell_i\})}
\bigg) = 0\;,
\end{align}
which can be recast as linear relations among integrals with shifted exponents,
\begin{align}
\sum_i\mu_i I(\sigma_1+\rho_{i,1},\dots,\sigma_n+\rho_{i,n}) = 0
\end{align}
for $\rho_{i,j}\in\{-1,0,1\}$. The virtue of IBP relations is that within a
given topology, a few integrals may be chosen as masters in the sense that all
other integrals can be expressed in a basis of them. The importance is not to
be underestimated. For example, the four-point massless planar triple box has 
several hundred renormalizable integrals which are reduced onto a linear
combination of just three integrals with at most rank 1.

IBP relations can be generated by public computer codes such as {\tt FIRE}
\cite{Smirnov:2013dia} and {\tt Reduze} \cite{vonManteuffel:2012np}. In
practice, the production of IBP relations is quite time consuming and requires
considerable amount of memory. 

\subsection{Direct Extraction of Integral Coefficients}
We consider schematically an $n$-loop amplitude contribution which is denoted
$\mc A^{(L)}$. After reduction onto a basis of master integrals, the
amplitude can be written
\begin{align}
\label{MASTEREQ}
\mc A^{(L)} = 
\sum_{k\in\text{Basis}}c_kI_k+\text{rational terms}
\end{align}
where the $c_k$'s are rational functions of external invariants. We refer to
eq.~\eqref{MASTEREQ} as the {\it master equation}. For example, at one loop the
integral basis is very simple and contains only boxes, triangles, bubbles and
rational terms. The integral itself is calculated once and for all and
therefore the problem of computing the amplitude reduces to determining the
coefficients.

The trick is to probe the loop integrand by applying generalized unitarity
cuts on either side of the master equation. Originally, unitarity cuts were
realized by replacing a set of propagators by Dirac delta functions
restricting them to their mass shell. The framework of maximal unitarity at
two loops initiated by Kosower and Larsen \cite{Kosower:2011ty} naturally
deals with amplitude contributions whose factorization properties are
accessible only away from the real slices of Minkowski space, for example
hepta-cuts of double boxes. Multidimensional complex contour integrals that
compute multivariate residues provide the desired generalization of the
localization property,
\begin{align}
\int dz_1\cdots\int dz_n h(\{z_i\})\prod_{j=1}^n\delta(z_j-\xi_j)
\equiv
\frac{1}{(2\pi i)^n}\oint_{\Gamma_\epsilon(\xi)}\!
dz_1\wedge\cdots \wedge dz_n\frac{h(\{z_i\})}{\prod_{j=1}^n(z_j-\xi_j)}
\end{align}
for a given $\xi\in\C^n$. Here, $\Gamma_\epsilon$ is a torus of real dimension
$n$ around the pole of the integrand at $z = \xi$. A generalized
unitarity cut, even a maximal cut which puts as many propagators on-shell as
possible, is typically shared among several basis integrals, hence
intermediate algebra is in principle required. Instead one seeks to construct
linear combinations of residues that in a certain sense are orthogonal to each
other and thus project a single basis integral. In this way, the integral
coefficient is expressed in terms of residues of products of tree amplitudes
that arise when the loop amplitude factorizes. These combinations are subject to
the consistency requirement that parity-odd integrands and total derivatives
continue to vanish upon integration \cite{Kosower:2011ty}. The tree-level data
is easily manipulated within the spinor-helicity formalism by means of for
instance superspace techniques \cite{Sogaard:2011pr,Bern:2009xq}.

Direct extraction of master integral coefficients in maximal unitarity has
been demonstrated for two-loop double boxes with up to four external massive
or massless legs
\cite{Kosower:2011ty,Johansson:2012zv,Johansson:2013sda,CaronHuot:2012ab}, for
the nonplanar double box \cite{Sogaard:2013yga} and the three-loop triple box
\cite{Sogaard:2013fpa}. In these calculations, only basis integrals with
single propagators were considered. 

\section{Multivariate Residues}
To extract the integral coefficients, we need to calculate {\it multivariate
residues}. In many cases, these residues can simply be evaluated by Cauchy's
theorem in higher dimensions and the Jacobian determinant. However, in some
cases, like the unitarity cut of the triple box topology
\cite{Sogaard:2013fpa} or the unitarity cut of integral with doubled
propagators, the residues are {\it degenerate} and have to be evaluated by
algebraic geometry. In this section, we briefly review the concept and
calculation of multivariate residues. Standard mathematical references include
\cite{MR507725,MR0463157,MR2161985}. 

Consider a differential form $\omega$ in $n$ complex variables 
$z\equiv (z_1,\dots,z_n)$,
\begin{equation}
  \label{differential_form}
  \omega=\frac{h(z) dz_1 \wedge \cdots \wedge dz_n}{f_1(z) \cdots f_n(z)}\;,
\end{equation}
where the numerator $h(z)$ and the denominators $f_1(z), \dots, f_n(z)$ are
holomorphic functions. If at a point $\xi$, $f_1(\xi)=\cdots =f_n(\xi)=0$,
then the residue of $\omega$ at $\xi$ regarding the divisors 
$\{f_1,\dots,f_n\}$ is defined to be,
  \begin{equation}
    \label{local_residue}
    \Res{}_{\{f_1,
    \dots, f_n\},\xi}(\omega)\equiv \bigg(\frac{1}{2\pi i}\bigg)^n\oint_{\Gamma}
    \frac{h(z) dz_1 \wedge \cdots \wedge dz_n}{f_1(z) \cdots f_n(z)}\;.
  \end{equation}
Here the contour $\Gamma$ is a real $n$-cycle
$\Gamma=\{z: |f_i(z)|=\epsilon_i\}$ around $\xi$ and the orientation
is specified by $d(\arg f_1) \wedge \cdots \wedge d(\arg f_n)$.

In two cases, a multivariate residue can be calculated straightforwardly,
\begin{itemize}
\item A residue is {\it non-degenerate}, if the Jacobian at $\xi$ is
nonzero, i.e.,
\begin{equation} 
J(\xi)=\det(\partial f_i/\partial z_j)|_\xi\not=0\;.
\end{equation}
In this case, by the multi-dimensional verion of Cauchy's theorem, the value
of residue is simply \cite{MR507725},
\begin{equation}
    \label{non_degenerate_residue}
    \Res{}_{\{f_1,
    \dots, f_n\},\xi}(\omega)=\frac{h(\xi)}{J(\xi)}\;.
  \end{equation}
\item A residue is {\it factorizable } if each $f_i$ is a univariate
polynomial, namely, $f_i(z)=f_i(z_i)$. In this case, the $n$-dimensional
contour in is factorized to the product of $n$ univariate contours,
\begin{equation}
  \label{factorizable_residue}
   \Res{}_{\{f_1,
   \dots, f_n\},\xi}(\omega)=\bigg(\frac{1}{2\pi i}\bigg)^n 
   \oint_{|f_1(z_1)|=\epsilon_1}
   \frac{dz_1}{f_1(z_1)} \cdots
   \oint_{|f_n(z_n)|=\epsilon_n} \frac{dz_n}{f_n(z_n)} h(z)\;.
\end{equation}
Then, we can evaluate this residue by applying the univariate residue formula
$n$ times.
\end{itemize}

However, in general, a residue is neither non-degenerate nor factorizable. For
example, consider a Feynman integrand with doubled (or higher-power)
propagators,
\begin{equation}
  \label{eq:3}
  \frac{1}{f_1^{\sigma_1} f_2 \cdots f_n}\;, \quad \sigma_1 >1\;.
\end{equation}
At a point $\xi$ where $f_1(\xi)=\cdots =f_k(\xi)=0$, the Jacobian
matrix is degenerate since
\begin{equation}
  \label{eq:1}
  \frac{\partial f_1^{\sigma_1}}{\partial z_i} \bigg|_\xi=
  \sigma_1 f_1^{\sigma_1-1}  \frac{\partial
  f_1}{\partial z_i} \bigg|_\xi =0\;.
\end{equation}
In general, this type of residues is not factorizable. To evaluate
them, we need the {\it transformation law} \cite{MR507725} in algebraic geometry.

\begin{thm}[Transformation law] Let $\{f_1,\dots,f_n\}$ and
  $\{u_1,\dots,u_n\}$ be two sets of holomorphic functions and 
 $u_i = a_{ij} f_j$, where $a_{ij}$ are holomorphic functions. Assume that for
 each set, the common zeros are discrete points. Let $A$ be the
 matrix of the $a_{ij}$'s, then 
 \begin{equation}
   \label{transformation_law}
   \Res{}_{\{f_1,\dots,f_n\}, \xi}\bigg(\frac{h(z) dz_1 \wedge \cdots
     \wedge dz_n}{f_1(z) \cdots f_n(z)}\bigg) =  \Res{}_{\{u_1,\dots,
     u_n\}, \xi}   \bigg(\frac{h(z) dz_1 \wedge \cdots
     \wedge dz_n}{u_1(z) \cdots u_n(z)} \det A \bigg)\;.
 \end{equation}
\end{thm}
This theorem holds for both non-degenerate and degenerate residues.  

For Feynman integrals, the denominators are all polynomials. In this
case, we can use the transformation law to convert a degenerate
residue to a factorizable residue, via Gr\"obner basis. The algorithm
involves the following steps \cite{Sogaard:2013fpa}:
\begin{enumerate}
\item Calculate the Gr\"obner basis $\{g_1,\ldots,g_k\}$ of
$\{f_1,\ldots,f_n\}$ in the {\it DegreeLexicographic} order and record the
converting matrix
  $r_{ij}$, such that $g_i=r_{ij} f_j$.
\item For each $1\leq i \leq n$, calculate the Gr\"obner basis of
$\{f_1,\ldots,f_n\}$ in the {\it
    Lexicographic} order of $z_{i+1} \succ \cdots \succ z_n \succ z_1
  \succ \cdots z_i$. Pick the univariate polynomial in $z_i$ from this
  Gr\"obner basis and name it as $u_i$.  \item For each $u_i$, divide it towards
  $\{g_1,\ldots,g_k\}$ so
  $u_i=s_{ij} g_j$. 
\item The transformation matrix is $a_{ij}= s_{ik} r_{kj}$. By the transformation
  law, the degenerate residue is converted to a factorizable residue with the
  matrix $a_{ij}$.
\end{enumerate}

Finally, we have a comment on the residues from the maximal cut of
integrals with doubled (or multiple) propagators. For the residue from a
general integrand,
\begin{equation}
  \label{general_integrand}
  \frac{N}{f_1^{\sigma_1} \cdots f_n^{\sigma_n}}\;,
\end{equation}
to use (\ref{local_residue}), for each $f_i$, we have to collect all powers of
$f_i$ as one denominator. Otherwise, the common zeros of denominators are not
discrete points, so the residue is not well defined. Hence there is no
ambiguity of defining denominators for the residue computation.

In our paper, we calculate the residues from the maximal cut of integrals with
doubled and tripled propagators. The degenerate residues are evaluated by our
Mathematica package \texttt{MathematicaM2}\footnote{
{The package can be downloaded from \tt
https://bitbucket.org/yzhphy/mathematicam2}}, which calls Macaulay2 \cite{M2}
to finish the computation of Gr\"obner bases. We demonstrate the multivariate
residue computation explicitly by the one-loop box integral with double
propagators. Then we show this method is general by two-loop examples. Related
ideas were previously proposed to reduce one-loop integrals with generic
powers of propagators \cite{Zhang:2011ns}.

\section{Example: One-Loop Box}
Consider a one-loop box integral with four massless legs 
$k_1,\ldots,k_4$,
\begin{align}
I_4(\sigma_1,\dots,\sigma_4)[N]\equiv {} &
\int_{\R^D}\!\frac{d^D\ell}{(2\pi)^D}
\prod_{i=1}^4\frac{N}{f_i^{\sigma_i}(\ell)}\;,
\end{align}
where the denominators are,
\begin{align}
f_1 = {} \ell^2\;, \quad
f_2 = {} (\ell-k_1)^2\;, \quad
f_3 = {} (\ell-K_{12})^2\;, \quad
f_4 = {} (\ell+k_4)^2\;.
\end{align}
We suppress the Feynman $i\epsilon$-prescription as it is irrelevant for our
purposes and assume for simplicity that all external momenta are massless and
outgoing. Multiple consecutive external momenta are summed using the shorthand
notation $K_{ij} = k_i+\cdots+k_j$.

In the following discussion, we set $D=4$. We fix a basis of the
four-dimensional space time $\{k_1,k_2,k_4,\omega\}$ where the spurious vector
$\omega$ can be represented as
\begin{align}
\omega\equiv\frac{1}{2s_{12}}\left(
\spab{2}{3}{1}\spab{1}{\gamma^\mu}{2}-
\spab{1}{3}{2}\spab{2}{\gamma^\mu}{1}
\right)\;,
\end{align}
such that $\omega$ is orthogonal to the subspace spanned by the momentum
vectors. The list of irreducible scalar products (ISP) can then be chosen as
\begin{align}
\ISP = \{\ell\cdot\omega\}\;.
\end{align}

The loop momentum $\ell$ can be parameterized as 
\begin{align}
\label{PARAM}
\ell^\mu & = 
\alpha_1k_1^\mu+\alpha_2k_2^\mu+
\frac{s_{12}\alpha_3}{2\spaa{14}\spbb{42}}\expval{1^-}{\gamma^\mu}{2^-}+
\frac{s_{12}\alpha_4}{2\spaa{24}\spbb{41}}\expval{2^-}{\gamma^\mu}{1^-}\;,
\end{align}
and the Jacobian for this parametrization is
\begin{align}
J= \det_{\mu,i}\pd{\ell^\mu}{\alpha_i} =
-\frac{is_{12}^2}{4\chi(\chi+1)}\;.
\end{align}

The cut equations $f_1(\ell)=\cdots =f_4(\ell)=0$ have two solutions, 
\begin{eqnarray}
  \label{eq:2}
(\alpha_1,\ldots \alpha_4)  = (1,0,0,-\chi)\equiv \xi_1\;, \quad
(\alpha_1,\ldots \alpha_4)  = (1,0,-\chi, 0) \equiv \xi_2\;.
\end{eqnarray}
There is only one master integral for one-loop box, the scalar
integral,
\begin{equation}
  \label{eq:7}
  I_4(\sigma_1,\ldots \sigma_4)[N]=C_1 I_4(1,1,1,1)[1]_{\xi_1} + \cdots
\end{equation}
where $\cdots$ stands for integrals with fewer than four propagators.

Localizing the contour around $\xi_1$ and $\xi_2$, it is clear that the
Jacobian of $f$'s in $\alpha$'s is nonzero. So by Cauchy's theorem in higher
dimensions (\ref{non_degenerate_residue}), the residues are
\begin{equation}
  \label{eq:6}
  I_4(1,1,1,1)[1]_{\xi_1}=-i\frac{1}{4s_{12}^2\chi},\quad
I_4(1,1,1,1)[1]_{\xi_2}=i\frac{1}{4s_{12}^2\chi}\;.
\end{equation}
Together with the spurious integral condition $ I_4(1,1,1,1)[\ell\cdot
\omega]_{\xi_1}=0$, we have the expression for the integral
coefficient for the integral $I_4(\sigma_1,\ldots \sigma_4)[N]$,
\begin{equation}
  \label{eq:8}
  C_1 =  2 i s_{12}^2 \chi (I_4(\sigma_1,\ldots \sigma_4)[N]_{\xi_1} -
I_4(\sigma_1,\ldots \sigma_4)[N]_{\xi_2})\;.
\end{equation}

Now we consider integrals with doubled propagators, for example,
$I_4(1,1,1,2)[1]$. The Jacobian of $\{f_1, f_2 ,f_3, f_4^2\}$ in
$\alpha$'s is zero at both $\xi_1$ and $\xi_2$, so direct computation
does not work. We can use the transformation law \eqref{transformation_law} to
convert the denominators to a calculable form,
\begin{equation}
  \label{eq:9}
\left(
  \begin{array}{c}
 g_1\\
g_2\\
g_3\\
g_4
  \end{array}
\right)\equiv
\left(
  \begin{array}{c}
    \alpha_1-1\\
\alpha_2\\
-\alpha_3(\alpha_3+\chi)^2\\
-\alpha_4(\alpha_4+\chi)^2\\
  \end{array}
\right)
=M 
\left(
\begin{array}{c}
 f_1\\
f_2\\
f_3\\
f_4^2
  \end{array}
\right)\;,
\end{equation}
where $M$ is $4\times 4$ matrix and all entries are polynomials in
$\alpha$'s, and
\begin{equation}
  \label{eq:10}
  \det M=-\frac{\chi(1+\chi)
  (\alpha_3-\alpha_4)(\alpha_3+\alpha_4+2\chi)}{s_{12}^5}\;.
\end{equation}
Hence, around either $\xi_1$ or $\xi_2$,
\begin{equation}
  \label{eq:11}
  \oint \frac{d\alpha_1\wedge d\alpha_2\wedge d\alpha_3\wedge
    d\alpha_4}{f_1 f_2 f_3 f_4^2} =  \oint
  \frac{d\alpha_1}{\alpha_1-1} \oint \frac{d\alpha_2}{\alpha_2}
  \oint\frac{d\alpha_3}{\alpha_3(\alpha_3+\chi)^2}\oint
    \frac{\det M  d\alpha_4 }{\alpha_4(\alpha_4+\chi)^2}\;.
\end{equation}
So the degenerate residue can be calculated by applying the univariate
Cauchy's theorem four times. The explicit form of $M$ is found by our package
\texttt{MathematicaM2}. The residues for $I_4(1,1,1,2)[1]$ are
\begin{equation}
  I_4(1,1,1,2)[1]_{\xi_1}=-i\frac{1}{4s_{12}^3\chi^2}\;,\quad
  I_4(1,1,1,2)[1]_{\xi_2}=i\frac{1}{4s_{12}^3\chi^2}\;.
\end{equation}
So we have
\begin{equation}
  \label{eq:12}
   I_4(1,1,1,2)[1]=\frac{1}{s_{12} \chi}I_4(1,1,1,1)[1] + \cdots\;.
\end{equation}
Similarly, using the same method, we find that,
\begin{eqnarray}
  \label{eq:13}
     I_4(2,1,1,1)[1]&=&\frac{1}{s_{12} }I_4(1,1,1,1)[1] + \cdots\;, \\
     I_4(2,1,1,2)[1]&=&\frac{2}{s_{12}^2\chi }I_4(1,1,1,1)[1] + \cdots\;, \\
     I_4(3,1,1,1)[1]&=&\frac{1}{s_{12}^2 }I_4(1,1,1,1)[1] + \cdots\;.
\end{eqnarray}

These results are consistent with the IBP relations in the $D=4$
limit. For instance, from {\tt FIRE} \cite{Smirnov:2013dia}, 
\begin{eqnarray}
  \label{eq:14}
    I_4(1,1,1,2)[1]&=&\frac{1+2\epsilon}{s_{12} \chi}I_4(1,1,1,1)[1] +\cdots\;,\\
     I_4(2,1,1,1)[1]&=&\frac{1+2\epsilon}{s_{12} }I_4(1,1,1,1)[1] + \cdots\;, \\
     I_4(2,1,1,2)[1]&=&\frac{2(1+\epsilon)(1+2\epsilon)}{s_{12}^2\chi}I_4(1,1,1,1)[1] + \cdots\;, \\
     I_4(3,1,1,1)[1]&=&\frac{(1+\epsilon)(1+2\epsilon)}{s_{12}^2}I_4(1,1,1,1)[1] + \cdots\;.
\end{eqnarray}

\section{Example: Planar Double Box}
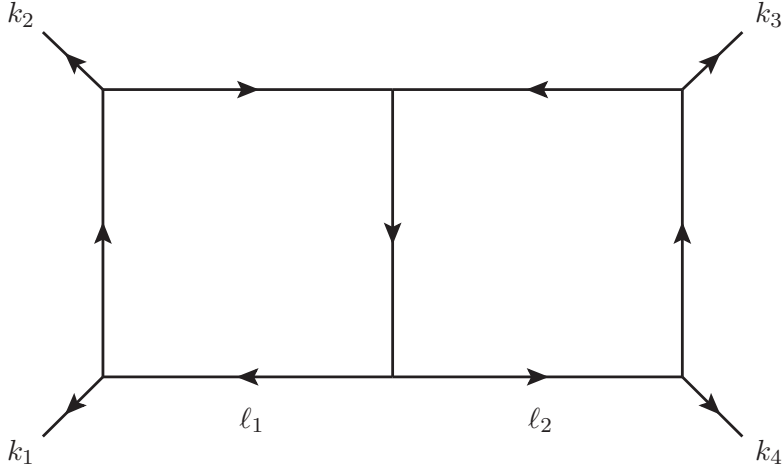
\begin{figure}[!h]
\bc
\includegraphics[scale=0.8]{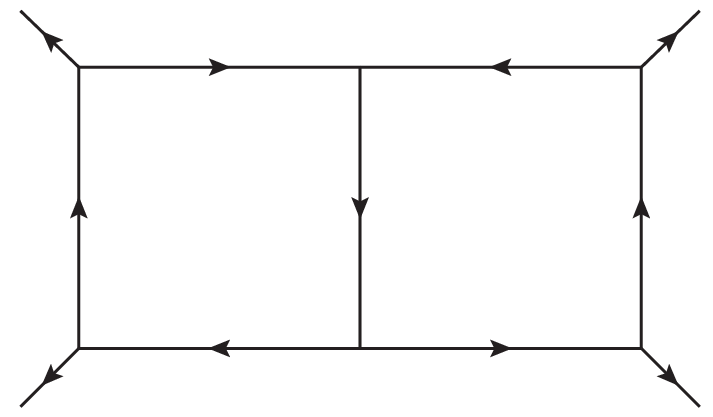}
\put(-280,-5){$k_1$}
\put(-280,160){$k_2$}
\put(0,160){$k_3$}
\put(0,-5){$k_4$}
\put(-193,7){$\ell_1$}
\put(-85,7){$\ell_2$}
\caption{The massless four-point planar double box.}
\label{DBOX}
\ec
\end{figure}

We now proceed to two-loop integrals. The generalized dimensionally
regularized two-loop planar double box scalar integral (fig.~\ref{DBOX}) with
arbitrary powers of propagators reads
\begin{align}
\P(\sigma_1,\dots,\sigma_7) \equiv {} &
\int_{\R^D}\!\frac{d^D\ell_1}{(2\pi)^D}
\int_{\R^D}\!\frac{d^D\ell_2}{(2\pi)^D}
\prod_{i=1}^7\frac{1}{f_i^{\sigma_i}(\ell_1,\ell_2)}\;,
\end{align}
where the seven inverse propagators $\{f_i\}$ are given by
\begin{align}
f_1 = {} & \ell_1^2\;, &
f_2 = {} & (\ell_1-k_1)^2\;, &
f_3 = {} & (\ell_1-K_{12})^2\;, \nn \\
f_4 = {} & \ell_2^2\;, &
f_5 = {} & (\ell_2-k_4)^2\;, & 
f_6 = {} & (\ell_2-K_{34})^2\;, &
f_7 = {} & (\ell_1+\ell_2)^2\;. &
\end{align}
Closed form expressions for planar double integrals can be found in
refs.~\cite{Smirnov:1999gc,Smirnov:1999wz}.

As in the one-loop example, we choose $\{k_1,k_2,k_4,\omega\}$ as basis of the
four-dimensional space time where again the spurious vector $\omega$ can be
represented as
\begin{align}
\omega\equiv\frac{1}{2s_{12}}\left(
\spab{2}{3}{1}\spab{1}{\gamma^\mu}{2}-
\spab{1}{3}{2}\spab{2}{\gamma^\mu}{1}
\right)\;,
\end{align}
The list of ISPs can then be chosen as
\cite{Badger:2012dp}
\begin{align}
\ISP = \{\ell_1\cdot k_4,\,\ell_2\cdot k_1,
\,\ell_1\cdot\omega,\,\ell_2\cdot\omega\}\;,
\end{align}
and the integrand basis contains 16 spurious and 16 nonspurious elements.
Whence the nine-propagator double box topology is defined by
\begin{align}
\P(\sigma_1,\dots,\sigma_9) \equiv {} &
\int_{\R^D}\!\frac{d^D\ell_1}{(2\pi)^D}
\int_{\R^D}\!\frac{d^D\ell_2}{(2\pi)^D}
\prod_{i=1}^9\frac{1}{f_i^{\sigma_i}(\ell_1,\ell_2)}\;, 
\end{align}
where $f_8 = \ell_1\cdot k_4$ and $f_9 = \ell_2\cdot k_1$ are the nonspurious
ISPs. Then we have
\begin{align}
\P[(\ell_1\cdot k_4)^n(\ell_1\cdot k_2)^m] = 
\P(1,\dots,1,-n,-m)
\end{align}
in the notation of refs.~\cite{Gluza:2010ws,Kosower:2011ty}.

\subsection{Parametrization of Hepta-Cut Solutions}
In order to expose the singularity structure of the loop integrand, we adopt a
particularly convenient loop momentum parametrization of previous works, see
for instance ref.~\cite{Kosower:2011ty},
\begin{align}
\label{LOOPPARAM}
\ell_1^\mu & = 
\alpha_1k_1^\mu+\alpha_2k_2^\mu+
\frac{s_{12}\alpha_3}{2\spaa{14}\spbb{42}}\expval{1^-}{\gamma^\mu}{2^-}+
\frac{s_{12}\alpha_4}{2\spaa{24}\spbb{41}}\expval{2^-}{\gamma^\mu}{1^-}\;,
\\[1mm]
\ell_2^\mu & = 
\beta_1k_3^\mu+\beta_2k_4^\mu+
\frac{s_{12}\beta_3}{2\spaa{31}\spbb{14}}\expval{3^-}{\gamma^\mu}{4^-}+
\frac{s_{12}\beta_4}{2\spaa{41}\spbb{13}}\expval{4^-}{\gamma^\mu}{3^-}\;.
\end{align}
It is elementary to show that the Jacobians to compensate for the change of
variables from loop momenta to parameter space are
\begin{align}
\label{LOOPPARAMETERJACOBIAN}
J_\alpha = \det_{\mu,i}\pd{\ell_1^\mu}{\alpha_i} =
-\frac{is_{12}^2}{4\chi(\chi+1)}\;, \quad
J_\beta = \det_{\mu,i}\pd{\ell_2^\mu}{\beta_i} = 
-\frac{is_{12}^2}{4\chi(\chi+1)}\;,
\end{align}
where $\chi$ is a frequently used ratio of Mandelstam invariants,
\begin{align}
\chi = \frac{s_{14}}{s_{12}}\;.
\end{align}
The zero locus of the ideal generated by the polynomials $f_i$ defines a
reducible elliptic curve associated with a hextuply pinched torus whose
components are Riemann spheres,
\begin{align}
\mc S = 
\big\{(\ell_1,\ell_2)\in(\C^4)^{\otimes 2}\;|\; 
f_i(\ell_1,\ell_2) = 0\big\} =
\mc S_1\cup\cdots\cup \mc S_6\;.
\end{align}
The solutions can be summarized as follows,
\begin{align}
\mc S_1\;& : (\alpha_3,\alpha_4,\beta_3,\beta_4) = (-\chi,0,z,0)\;, \quad &
\mc S_2\;& : (\alpha_3,\alpha_4,\beta_3,\beta_4) = (z,0,-\chi,0)\;, \\
\mc S_3\;& : (\alpha_3,\alpha_4,\beta_3,\beta_4) = (0,-\chi,0,z)\;, \quad &
\mc S_4\;& : (\alpha_3,\alpha_4,\beta_3,\beta_4) = (0,z,0,-\chi)\;, \\
\mc S_5\;& : (\alpha_3,\alpha_4,\beta_3,\beta_4) = (0,z,\tau(z),0)\;, \quad &
\mc S_6\;& : (\alpha_3,\alpha_4,\beta_3,\beta_4) = (z,0,0,\tau(z))\;,
\end{align}
with $(\alpha_1,\alpha_2,\beta_1,\beta_2) = (1,0,0,1)$ uniformly across all
branches. Also, $\tau$ is defined by
\begin{align}
\tau(z)\equiv -(\chi+1)\frac{z+\chi}{z+\chi+1}\;.
\end{align}

\subsection{Residues of the Loop Integrand}
We follow the strategy of refs.~\cite{Kosower:2011ty,CaronHuot:2012ab} and
quickly rederive the hepta-cut of the massless planar double box. In the
standard situation where all propagators are single, the residue of the scalar
integrand is nondegenerate and is easy to calculate as the determinant of a
Jacobian matrix as explained previously. For all six branches, the hepta-cut
integral is \cite{Kosower:2011ty}
\begin{align}
\label{DBOXCUT}
\P(1,\dots,1,0,0)_{\mc S_i} = 
-\frac{1}{16s_{12}^3}\oint\frac{dz}{z(z+\chi)}\;.
\end{align}
It remains to choose an integral basis. The IBP identities generated with {\tt
FIRE} \cite{Smirnov:2013dia} grant that all double box integrals can be
reduced onto two master integrals, such that a general integral (and the
amplitude contribution itself) can be written
\begin{align}
\label{DBOXMIEQ}
\P(\sigma_1,\dots,\sigma_9) = 
C_1P_{2,2}^{**}(1,\dots,1,1,0,0)+
C_2P_{2,2}^{**}(1,\dots,1,-1,0)+\cdots
\end{align}
where hidden terms have less than seven propagators and therefore vanish on
the maximal cut. Evidently, the integral basis consists of a scalar double box 
and a rank 1 tensor integral with single propagators.

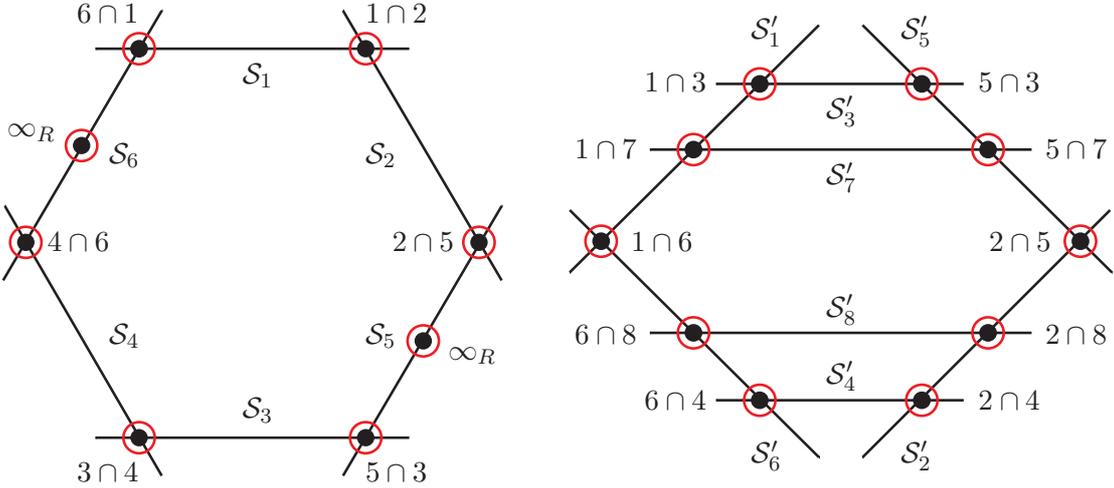
\begin{figure}[!h]
\bc
\includegraphics[scale=0.65]{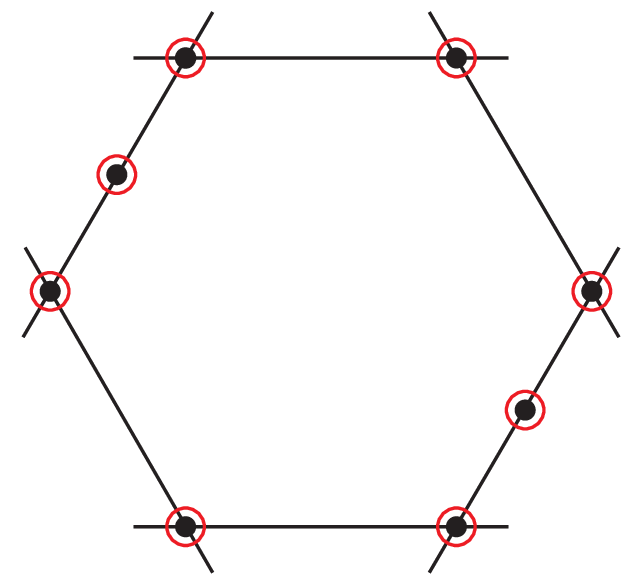} \;\;\;
\put(-115,152){$\mc S_1$}
\put(-115,26){$\mc S_3$}
\put(-165,54){$\mc S_4$}
\put(-69,54){$\mc S_5$}
\put(-69,122){$\mc S_2$}
\put(-165,122){$\mc S_6$}
\put(-203,132){$\infty_R$}
\put(-38,48){$\infty_R$}
\put(-59,89){$2\cap 5$}
\put(-188,89){$4\cap 6$}
\put(-177,2){$3\cap 4$}
\put(-177,175){$6\cap 1$}
\put(-69,2){$5\cap 3$}
\put(-69,175){$1\cap 2$}
\raisebox{1.4mm}{
\includegraphics[scale=0.65,angle=90]{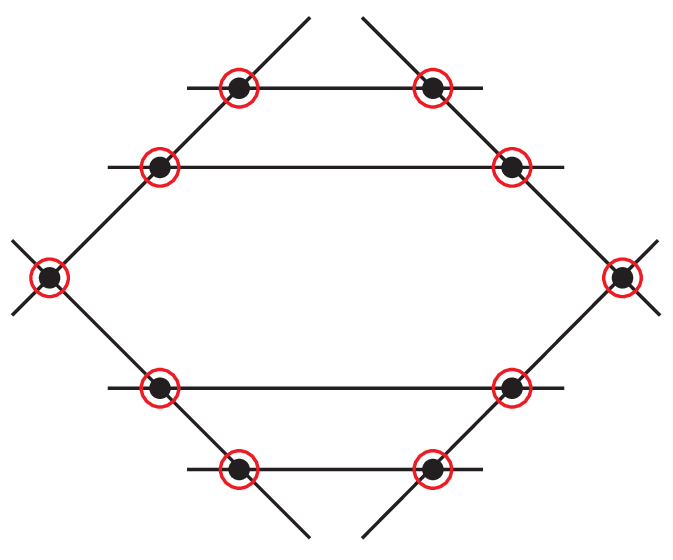}
\put(-111,35){$\mc S'_4$}
\put(-111,60){$\mc S'_8$}
\put(-111,110){$\mc S'_7$}
\put(-111,135){$\mc S'_3$}
\put(-139,5){$\mc S'_6$}
\put(-83,5){$\mc S'_2$}
\put(-139,165){$\mc S'_1$}
\put(-83,165){$\mc S'_5$}
\put(-54,145){$5\cap 3$}
\put(-29,120){$5\cap 7$}
\put(-179,145){$1\cap 3$}
\put(-205,120){$1\cap 7$}
\put(-205,50){$6\cap 8$}
\put(-179,25){$6\cap 4$}
\put(-54,25){$2\cap 4$}
\put(-29,50){$2\cap 8$}
\put(-184,85){$1\cap 6$}
\put(-50,85){$2\cap 5$}
}
\caption{Global structure of the hepta-cut of the two-loop planar (left) and
nonplanar (right) double box with purely massless kinematics and four external
legs. The straight lines should be interpreted as genus-0 Riemann surfaces.
Each branch may have an additional residue at $z = \infty$ which is eliminated
here.}
\label{DBOXXBOX_GLOBAL}
\ec
\end{figure}

We will focus on integrals that have at least one $\sigma_i > 1$, e.g.
\begin{align}
\label{DBOXDOUBLEDEXAMPLES}
\P(2,1,\dots,1,0,0) = 
C_1\P(1,\dots,1,0,0)+C_2\P(1,\dots,1,-1,0)+\cdots\; \\[1mm]
\P(1,\dots,1,2,0,0) = 
C_1'\P(1,\dots,1,0,0)+C_2'\P(1,\dots,1,-1,0)+\cdots\;
\end{align}
for rational coefficients in external invariants and space-time dimension.
The residue at the simultaneous zero of the inverse propagators in such an
integral is clearly degenerate by the preceding discussion. Therefore we
apply the algorithm to transform the residue to a factorized form and obtain
the analog of eq.~\eqref{DBOXCUT} 
\begin{align}
\P(2,1,\dots,1,0,0)_{\mc S_{1,3}} = {} &
-\frac{1}{16s_{12}^4}\oint\frac{dz}{z(z+\chi)}\;, \\[1mm]
\P(2,1,\dots,1,0,0)_{\mc S_{2,4,5,6}} = {} &
-\frac{1}{16s_{12}^4}\oint\frac{dz}{z(z+\chi)^2}\;, \\[1mm]
\P(1,\dots,1,2,0,0)_{\mc S_i} = {} &
-\frac{1}{16s_{12}^4}\oint\frac{dz}{z(z+\chi)^2}\;.
\end{align}
The integrand thus has two residues at each branch (we eliminate residues at
infinity by the Global Residue Theorem) and since solution $\mc S_5$ and $\mc
S_6$ give rise to additional poles in numerator insertions, we expect a total
of fourteen independent residues. However, as explained in 
refs.~\cite{CaronHuot:2012ab,Huang:2013kh} and depicted in
fig.~\ref{DBOXXBOX_GLOBAL}, the Jacobian poles are located at the nodal points
of the elliptic curve defined by the hepta-cut. We truncate to a linearly
independent set of residues for $\mc S_1\cup\cdots\cup S_6$ and choose
contours for the post hepta-cut degree of freedom that encircle eight global,
\begin{align}
(\mc G_1,\dots,\mc G_8) = 
(\mc G_{1\cap2},\,\mc G_{2\cap5},\,\mc G_{5\cap3}\,,
\mc G_{3\cap4},\,\mc G_{4\cap6},\,\mc G_{6\cap1},\,
\mc G_{5,\infty_R},\,\mc G_{6,\infty_R}) = \;,
\end{align}
with the corresponding weights,
\begin{align}
\Omega = (\omega_{1\cap2},\,\omega_{2\cap5},\,\omega_{5\cap3}\,,
\omega_{3\cap4},\,\omega_{4\cap6},\,\omega_{6\cap1},\,
\omega_{5,\infty_R},\,\omega_{6,\infty_R})\;.
\label{WINDINGNUM}
\end{align}
By convention, a residue with weight $\omega_{i\cap j}$ is evaluated on the
$i$th branch. 

The associated residues of the integrand in the two master integrals in
eq.~\eqref{DBOXMIEQ} in the order displayed above then read
\begin{align}
R_1 = \frac{1}{16\chi s_{12}^3}(1,-1,1,1,-1,1,0,0)\;, \quad
R_2 = \frac{1}{32s_{12}^2}(0,-1,0,0,-1,0,0,0)\;.
\end{align}

\subsection{Master Integral Projectors}
The hepta-cut contours are subject to consistency requirements in order to
ensure that certain integral relations are preserved after pushing the real
slice integrals into $\C^8$. It is well known that the integrand can be
parametrized in terms of four irreducible products,
\begin{align}
N = \sum_{a_1,\dots,a_4}c_{a_1,\dots,a_4}
(\ell_1\cdot\omega)^{a_1}
(\ell_2\cdot\omega)^{a_2}
(\ell_1\cdot k_4)^{a_3}
(\ell_2\cdot k_1)^{a_4}\;,
\end{align}
whose powers can be derived by renormalizability conditions and then
multivariate polynomial division using the Gr\"obner basis method. The latter
has been automated in the program BasisDet \cite{Zhang:2012ce}. Both the
spurious and nonspurious part of the basis contains sixteen elements. At the
level of integrated expressions, the amplitude is expressed as a linear
combination of just two masters. Accordingly, we demand that all integral
identities on which the reduction relies are respected. The full list of IBP
identities is available in ref.~\cite{Sogaard:2013yga}. We arrange all
parity-odd and IBP constraints as a $32\times 8$ matrix $M$ that acts on the
residue weights $\Omega$. The two corresponding submatrices have rank 4 and 2
respectively. We then define two master contours by 
\begin{align}
\mc M_1\cdot(R_1,R_2) = (1,0)\;, \quad
\mc M_2\cdot(R_1,R_2) = (0,1)\;.
\end{align}
Here, $\mc M_1$ and $\mc M_2$ are particular choices of the winding numbers
\eqref{WINDINGNUM}, such that only the contribution from one of the basis
integrals is picked up and normalized. The full $34\times 8$ residue matrix
has rank 8 in either case and therefore the master contours are unique. The
solutions take the very simple form
\begin{align}
\mc M_1 = 4\chi s_{12}^3(1,0,1,1,0,1,1,1)\;, \quad
\mc M_2 = -8s_{12}^2(1,2,1,1,2,1,3,3)\;.
\end{align}

Now we are ready to apply the master integral projectors in the context of
double box integrals with doubled propagators. The residues of the integrands
of the integrals on the left hand side of eq.~\eqref{DBOXDOUBLEDEXAMPLES} are
\begin{align}
\Res{}_{\{\mc G_i\}}
\P(2,1,\dots,1,0,0) = {} & \frac{1}{16\chi s_{12}^4}(1,-1,1,1,-1,1,0,0)\;, 
\nn \\
\Res{}_{\{\mc G_i\}}
\P(1,\dots,1,2,0,0) = {} & \frac{1}{16\chi^2 s_{12}^4}(1,-1,1,1,-1,1,0,0)\;.
\end{align}
Therefore, applying the projectors yields the reduction identities
\begin{align}
\P(2,1,\dots,1,0,0) = {} & 
+\frac{1}{s_{12}}\P(1,\dots,1,0,0)+\cdots\;, \\
\P(1,\dots,1,2,0,0) = {} &
+\frac{1}{\chi s_{12}}\P(1,\dots,1,0,0)+\cdots\;.
\end{align}

It has been verified that our results are consistent with the four-dimensional
limit of the following IBP relations in $D = 4-2\epsilon$,
\begin{align}
\label{DBOXIBP}
\P(2,1,\dots,1,0,0) = {} &
\frac{1+2\epsilon}{s_{12}}\P(1,\dots,1,0,0)+\cdots \\[1mm]
\P(1,\dots,1,2,0,0) = {} &
\frac{1+2\epsilon}{1+\epsilon}\!\left(
\frac{1+3\epsilon}{\chi s_{12}}P_{2,2}^{**}(1,\dots,1,0,0)+
\frac{4\epsilon}{\chi s_{12}^2}P_{2,2}^{**}(1,\dots,1,-1,0)\right)+\cdots
\end{align}
Indeed, as $\epsilon\to 0$ the tensor integral drops out and the integral with
a doubled propagator and the canonical scalar integrals equate up the factors
written above.

Any other powers of propagators may be treated similarly. A complete list of
hepta-cuts of planar doubled boxes with a doubled propagator is given in
appendix \ref{DBOXDOUBLEDCUTS}.

\section{Example: Nonplanar Double Box}
\begin{figure}[!h]
\bc
\includegraphics[scale=0.8]{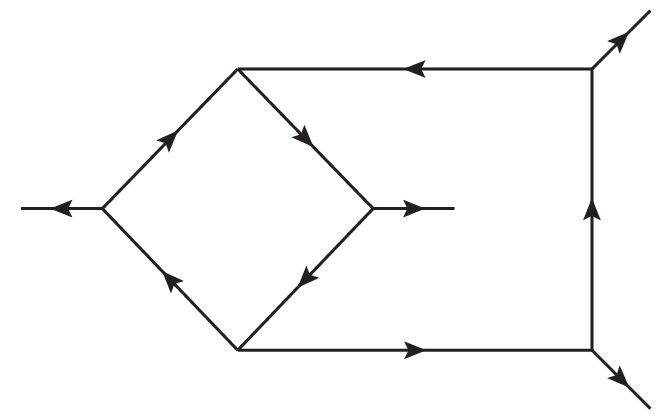}
\put(-261,78){$k_1$}
\put(-78,78){$k_2$}
\put(0,163){$k_3$}
\put(0,-8){$k_4$}
\put(-205,113){$\ell_1$}
\put(-20,78){$\ell_2$}
\caption{The nonplanar double box topology with four external particles.}
\label{XBOXDIAGRAM}
\ec
\end{figure}
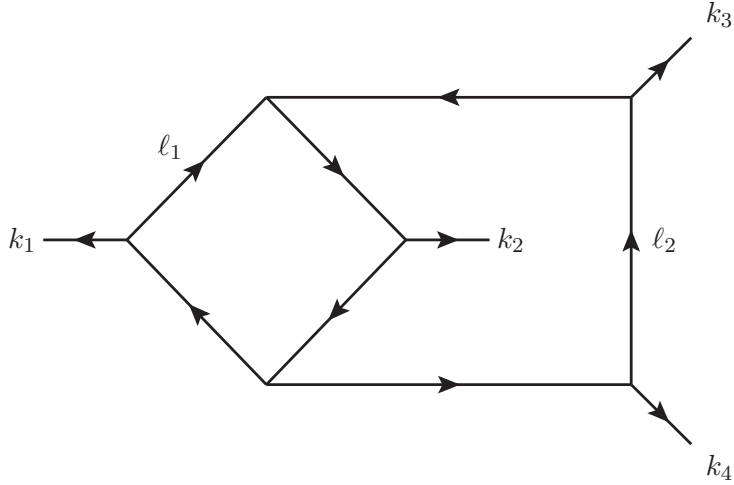

We define the four-point two-loop nonplanar double box integral
(see fig.~\ref{XBOXDIAGRAM}) in dimensional regularization by
\begin{align}
\X(\sigma_1,\dots,\sigma_7) \equiv {} &
\int_{\R^D}\!\frac{d^D\ell_1}{(2\pi)^D}
\int_{\R^D}\!\frac{d^D\ell_2}{(2\pi)^D}
\prod_{i=1}^7\frac{1}{\tilde f_i^{\sigma_i}(\ell_1,\ell_2)}\;,
\end{align}
and adopt the convention for propagators and momentum flow of
ref~\cite{Badger:2012dp},
\begin{align}
\tilde f_1 = {} & \ell_1^2\;, & \;
\tilde f_2 = {} & (\ell_1+k_1)^2\;, & \;
\tilde f_3 = {} & (\ell_2+k_4)^2\;, \nn \\
\tilde f_4 = {} & \ell_2^2\;, & \;
\tilde f_5 = {} & (\ell_1-k_3)^2\;, & \; 
\tilde f_6 = {} & (\ell_1+\ell_2-k_3)^2\;, & \; 
\tilde f_7 = {} & (\ell_1+\ell_2-K_{23})^2\;. &
\end{align}
All external and internal momenta are by assumption massless. We will consider
four-dimensional unitarity cuts and therefore only reconstruct the master
integral coefficients to leading order in the dimensional regulator. The
Feynman integral itself was calculated in
refs.~\cite{Tausk:1999vh,Anastasiou:2000mf}.

The set of vectors $\{k_1,k_2,k_3,\omega\}$ where $\omega$ is the spurious
direction forms a basis of four-dimensional momentum space. There are again
four irreducible scalar products,
\begin{align}
\ISP = \{\ell_1\cdot k_3,\,\ell_2\cdot k_2,
\,\ell_1\cdot\omega,\,\ell_2\cdot\omega\}\;,
\end{align}
and the minimal representation of the integrand consists of 19 spurious and 19
nonspurious monomials. Accordingly, we define the nine-propagator version of
the two-loop crossed box integral by
\begin{align}
\X(\sigma_1,\dots,\sigma_9) \equiv {} &
\int_{\R^D}\!\frac{d^D\ell_1}{(2\pi)^D}
\int_{\R^D}\!\frac{d^D\ell_2}{(2\pi)^D}
\prod_{i=1}^9\frac{1}{\tilde f_i^{\sigma_i}(\ell_1,\ell_2)}\;,
\end{align}
for $\tilde f_8 = \ell_1\cdot k_3$ and $\tilde f_9 = \ell_1\cdot k_2$.

\subsection{Parametrization of Hepta-Cut Solutions}
In the nonplanar case, the zero locus of the ideal generated by inverse
propagators defines a reducible genus-3 algebraic curve. The global structure
of the hepta-cut for any configuration of external legs and masses was
previously uncovered by computational algebraic geometry \cite{Huang:2013kh}.
In the purely massless limit, the zero locus decomposes into a union of eight
components which are exactly the inequivalent hepta-cut solutions,
\begin{align}
\tilde{\mc S} = 
\big\{(\ell_1,\ell_2)\in(\C^4)^{\otimes 2}\;|\; 
\tilde f_i(\ell_1,\ell_2) = 0\big\} =
\tilde{\mc S}_1\cup\cdots\cup \tilde{\mc S}_8\;.
\end{align}
The spinorial loop momentum parametrization \eqref{LOOPPARAM} applies equally
well to the nonplanar double box. However, we choose a slightly different
normalization in $\ell_2$ to adjust the flow direction in comparison with
refs.~\cite{Badger:2012dp,Sogaard:2013yga},
\begin{align}
\ell_1^\mu & = 
\alpha_1k_1^\mu+\alpha_2k_2^\mu+
\frac{s_{12}\alpha_3}{2\spaa{14}\spbb{42}}\expval{1^-}{\gamma^\mu}{2^-}+
\frac{s_{12}\alpha_4}{2\spaa{24}\spbb{41}}\expval{2^-}{\gamma^\mu}{1^-}\;,
\\[1mm]
\ell_2^\mu & = 
\beta_1k_3^\mu+\beta_2k_4^\mu+
\frac{s_{12}\beta_3}{2\spaa{32}\spbb{24}}\expval{3^-}{\gamma^\mu}{4^-}+
\frac{s_{12}\beta_4}{2\spaa{42}\spbb{23}}\expval{4^-}{\gamma^\mu}{3^-}\;.
\end{align}
The hepta-cut equations were solved using this parametrization in
refs.~\cite{Badger:2012dp,Sogaard:2013yga} and the resulting eight solutions
are quoted here in table \ref{XBOXSOLUTIONS}.
\begin{table}
\begin{align}
\begin{array}{@{}ccccccccc@{}}
\toprule
\;&\; \alpha_1 \;&\; \alpha_2 \;&\; \alpha_3 \;&\; \alpha_4 
\;&\; \beta_1 \;&\; \beta_2 \;&\; \beta_3 \;&\; \beta_4 \\  
\midrule
\mc S_1 \;&\; \chi-z \;&\; 0 \;&\; \chi(z-\chi-1) \;&\; 0
\;&\; 0 \;&\; 0 \;&\; z \;&\; 0 \\
\mc S_2 \;&\; \chi-z \;&\; 0 \;&\; 0 \;&\; \chi(z-\chi-1) 
\;&\; 0 \;&\; 0 \;&\; 0 \;&\; z \\
\mc S_3 \;&\; 0 \;&\; 0 \;&\; z \;&\; 0
\;&\; 0 \;&\; 0 \;&\; \chi \;&\; 0 \\
\mc S_4 \;&\; 0 \;&\; 0 \;&\; 0 \;&\; z
\;&\; 0 \;&\; 0 \;&\; 0 \;&\; \chi \\
\mc S_5 \;&\; \chi-z \;&\; 0 \;&\; 0 \;&\; (\chi+1)(z-\chi)
\;&\; 0 \;&\; 0 \;&\; z \;&\; 0 \\
\mc S_6 \;&\; \chi-z \;&\; 0 \;&\; (\chi+1)(z-\chi) \;&\; 0
\;&\; 0 \;&\; 0 \;&\; 0 \;&\; z \\
\mc S_7 \;&\; -1 \;&\; 0 \;&\; 0 \;&\; z
\;&\; 0 \;&\; 0 \;&\; 1+\chi \;&\; 0 \\
\mc S_8 \;&\; -1 \;&\; 0 \;&\; z \;&\; 0
\;&\; 0 \;&\; 0 \;&\; 0 \;&\; 1+\chi \\
\bottomrule
\end{array}
\nn
\end{align}
\vspace*{-.5cm}
\caption{Local hepta-cut solutions of the massless four-point nonplanar double
box.\label{XBOXSOLUTIONS}}
\end{table}

\subsection{Residues of the Loop Integrand}
Once the seven inverse propagators have been expanded in the loop momentum
parametrization, it is straightforward to derive the hepta-cuts of the
nonplanar double box scalar integral with single propagators as nondegenerate
multivariate residues \cite{Sogaard:2013yga},
\begin{align}
\X(1,\dots,1,0,0)_{\mc S_{3,4}} = {} & 
-\frac{1}{16s_{12}^3}\oint\frac{dz}{z(z+\chi)}\;, \\[1mm]
\X(1,\dots,1,0,0)_{\mc S_{7,8}} = {} &
-\frac{1}{16s_{12}^3}\oint\frac{dz}{z(z-\chi-1)}\;, \\[1mm]
\X(1,\dots,1,0,0)_{\mc S_{1,2,5,6}} = {} &
-\frac{1}{16s_{12}^3}\oint\frac{dz}{z(z-\chi)(z-\chi-1)}\;.
\end{align}

For the topology and kinematical configuration in consideration, there are two
master integrals, for instance the scalar integral and a rank 1 tensor.
Therefore a general integral can be written
\begin{align}
\X(\sigma_1,\dots,\sigma_9) = 
C_1\X(1,\dots,1,0,0)+C_2\X(1,\dots,1,-1,0)+\cdots\;.
\end{align}
We will consider integrals with doubled and also tripled propagators,
\begin{align}
\X(2,1,\dots,1,0,0) = 
C_1\X(1,\dots,1,0,0)+C_2\X(1,\dots,1,-1,0)+\cdots\;, \\[1mm]
\X(1,\dots,1,3,0,0) = 
C_1'\X(1,\dots,1,0,0)+C_2'\X(1,\dots,1,-1,0)+\cdots\;,
\end{align}
and reconstruct the coefficients in strictly four dimensions. Evaluating the
hepta-cuts of the displayed integrals by means of degenerate multivariate
residues yields
\begin{align}
\X(2,1,\dots,1,0,0)_{\mc S_{3,4}} = {} & 
+\frac{1}{16s_{12}^4}\oint dz\frac{1+(1+\chi)z}{z^2(z+\chi)^2}\;, \\[1mm]
\X(2,1,\dots,1,0,0)_{\mc S_{7,8}} = {} &
+\frac{1}{16s_{12}^4}\oint dz\frac{1+\chi}{z^2(z-\chi-1)}\;, \\[1mm]
\X(2,1,\dots,1,0,0)_{\mc S_{1,2,5,6}} = {} &
+\frac{1}{16s_{12}^4}\oint dz\frac{1}{z(z-\chi)^2(z-\chi-1)}\;,
\end{align}
\begin{align}
\X(1,\dots,1,3,0,0)_{\mc S_{3,4}} = {} & 
-\frac{1}{16s_{12}^5}\oint dz\frac{h(z)}{z(z+\chi)^5}\;, \\[1mm]
\X(1,\dots,1,3,0,0)_{\mc S_{7,8}} = {} &
-\frac{1}{16s_{12}^5}\oint dz\frac{(1+\chi)^2}{z(z-\chi-1)^3}\;, \\[1mm]
\X(1,\dots,1,3,0,0)_{\mc S_{1,2,5,6}} = {} &
-\frac{1}{16s_{12}^5}\oint dz\frac{1}{z(z-\chi)^3(z-\chi-1)}\;,
\end{align}
where the numerator function $h(z)$ is defined by
\begin{align}
h(z) = \chi^4-\chi^3(4z+1)+\chi^2(z(z+1)+1)+2\chi z(z+1)+z^2\;.
\end{align}
It is easiest to compare results with refs.~\cite{Sogaard:2013yga} if we
eliminate all residues at infinity by the Global Residue Theorem and thus only
encircle poles at the nodal points of the algebraic curve defined by the
hepta-cut. In this case, the parametrization is holomorphic and there are no
additional poles in tensor integrals. Referring to fig.~\ref{DBOXXBOX_GLOBAL},
the contour weights are
\begin{align}
\Omega = (
\omega_{1\cap6},\,\omega_{1\cap3},\,\omega_{1\cap7}\,,
\omega_{2\cap5},\,\omega_{2\cap4},\,\omega_{2\cap8},\,
\omega_{5\cap3},\,\omega_{5\cap7},\,\omega_{6\cap4},\,
\omega_{6\cap 8}
)\;.
\end{align}
The residues computed by the master integrals for contours in this ordering
read
\begin{align}
R_1 = {} &
\frac{1}{16\chi(1+\chi)s_{12}^3}
(-1,1+\chi,-\chi,-1,1+\chi,-\chi,1+\chi,-\chi,1+\chi,-\chi)\;, \\[1mm]
R_2 = {} &
\frac{1}{32s_{12}^2}(0,1,-1,0,1,-1,0,0,0,0)\;.
\end{align}
In advance of what follows, we also need to collect the residues of the
integrals in question with doubled and tripled propagators,
\begin{align}
\label{XBOXDOUBEX}
&\Res{}_{\{\mc G_i\}}\X(2,1,\dots,1,0,0) = \nn \\[-1mm]
&\qquad\qquad\frac{1}{16(1+\chi)\chi^2 s_{12}^4}
(-1,1-\chi^2,\chi^2,-1,1-\chi^2,\chi^2,1-\chi^2,\chi^2,1-\chi^2,\chi^2)
\;, \\[2mm]
\label{XBOXTRIPEX}
&\Res{}_{\{\mc G_i\}}\X(1,\dots,1,3,0,0) = \nn \\[-1mm]
&\qquad\qquad\frac{1}{16(1+\chi)\chi^3s_{12}^5}
(-1,1+\chi^3,-\chi^3,-1,1+\chi^3,-\chi^3,1+\chi^3,-\chi^3,1+\chi^3,-\chi^3)
\;.
\end{align}

\subsection{Master Integral Projectors}
By integrand-level reduction using {\tt BasisDet} \cite{Zhang:2012ce} we find
the general form of the nonplanar double box numerator, parametrized by the
four irreducible scalar products,
\begin{align}
N = \sum_{a_1,\dots,a_4}c_{a_1,\dots,a_4}
(\ell_1\cdot\omega)^{a_1}
(\ell_2\cdot\omega)^{a_2}
(\ell_1\cdot k_4)^{a_3}
(\ell_2\cdot k_1)^{a_4}\;.
\end{align}
The basis consists of 19 spurious and 19 nonspurious terms. Insisting that
the reduction onto the two master integrals is respected by the unitarity
procedure yields a $38\times 10$ matrix $M$ whose submatrices corresponding
to the parity-odd and parity-even parts are rank 5 and 3 respectively. Notice
that all IBP relations used in this calculation can be obtained from
ref.~\cite{Sogaard:2013yga}. The full residue matrix obtained by adding either
of the master integral projectors,
\begin{align}
\mc M_1\cdot(R_1,R_2) = (1,0)\;, \quad
\mc M_2\cdot(R_1,R_2) = (0,1)\;,
\end{align}
to $M$ has rank 10, which guarantees that the master contours are unique. In
detail, the projectors are characterized by the 10-tuples
\begin{align}
\mc M_1 = {} & 2\chi(1+\chi)s_{12}^3(-2,1,1,-2,1,1,1,1,1,1)\;, \\[2mm]
\mc M_2 = {} & 4s_{12}^2(
2(1+2\chi),1-2\chi,-3-2\chi,2(1+2\chi),1-2\chi, \nn \\ & \hspace*{1.5cm}
-3-2\chi,1-2\chi,-3-2\chi,1-2\chi,-3-2\chi)\;.
\end{align}
We can now take advantage of the projectors to extract the coefficients in
eq.~\eqref{XBOXDOUBEX},
\begin{align}
\X(2,1,\dots,1,0,0) = 
+\frac{1}{\chi s_{12}}\X(1,\dots,1,0,0)-\frac{4}{\chi s_{12}^2}
\X(1,\dots,1,-1,0)+\cdots
\end{align}
and similarly for the integral with a tripled propagator,
\begin{align}
\X(1\dots,1,3,0,0) = 
+\frac{1}{\chi^2 s_{12}^2}\X(1,\dots,1,0,0)
-\frac{4(1-\chi)}{\chi^2s_{12}^3}\X(1,\dots,1,-1,0)+\cdots
\end{align}

The validity of our predictions for the coefficients has been tested against
IBP relations generated by {\tt FIRE} \cite{Smirnov:2013dia}. Taking the 
$D = 4$ limit of the following identities,
\begin{align}
\label{XBOXIBP1}
\X(2,1,\dots,1,0,0) = {} &
+\frac{(1+2\epsilon)(1+(3+2\chi)\epsilon)}{(1+\epsilon)\chi s_{12}}
\X(1,\dots,1,0,0) \nn \\[2mm]
& \hspace*{1.5cm}
-\frac{4(1+2\epsilon)(1+4\epsilon)}{(1+\epsilon)\chi s_{12}^2}
\X(1,\dots,1,-1,0)+\cdots
\end{align}
\begin{align}
\label{XBOXIBP2}
&\X(1,\dots,1,3,0,0) = \nn \\[1mm]
&\qquad\qquad+\frac{(1+2\epsilon)(2+(9(1+\epsilon)+
2\chi(1+2\epsilon-2(1+\epsilon)\chi))\epsilon)}{(2+\epsilon)\chi^2 s_{12}^2}
\X(1,\dots,1,0,0) \nn \\[1mm] & \qquad\qquad\qquad
-\frac{4(1+2\epsilon)(1+4\epsilon)(2-2\chi(1+\epsilon)+3\epsilon)}
{(2+\epsilon)\chi^2 s_{12}^3}
\X(1,\dots,1,-1,0)+\cdots
\end{align}
shows that the results are consistent.

We also examined hepta-cuts of all other four-point nonplanar double box
integrals that have a doubled propagator. The results are similar to those
presented here. Refer to appendix~\ref{XBOXDOUBLEDCUTS} for a complete list.

\subsection{A Scalar Integral Basis}
\begin{figure}[!h]
\bc
\hspace*{-4mm}
\includegraphics[scale=0.75]{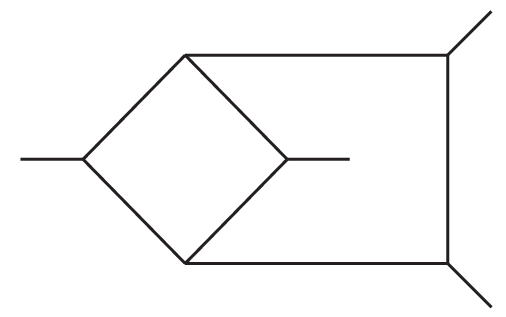} \hspace*{1cm}
\put(-221,55){$k_1$}
\put(-85,55){$k_2$}
\put(-34,112){$k_3$}
\put(-34,-3){$k_4$}
\put(-178,81){$\ell_1$}
\put(-47,55){$\ell_2$}
\includegraphics[scale=0.75]{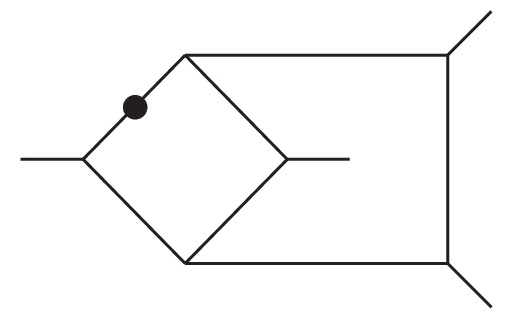}
\put(-190,55){$k_1$}
\put(-54,55){$k_2$}
\put(-3,112){$k_3$}
\put(-3,-3){$k_4$}
\put(-148,81){$\ell_1$}
\put(-16,55){$\ell_2$}
\caption{We use an integral basis for the massless four-point nonplanar double
box that contains no tensor numerators. Instead we have (left) a scalar
integral with single propagators and (right) a scalar integral with a doubled
propagator in the subbox.}
\label{XBOX_SUBBOX}
\ec
\end{figure}
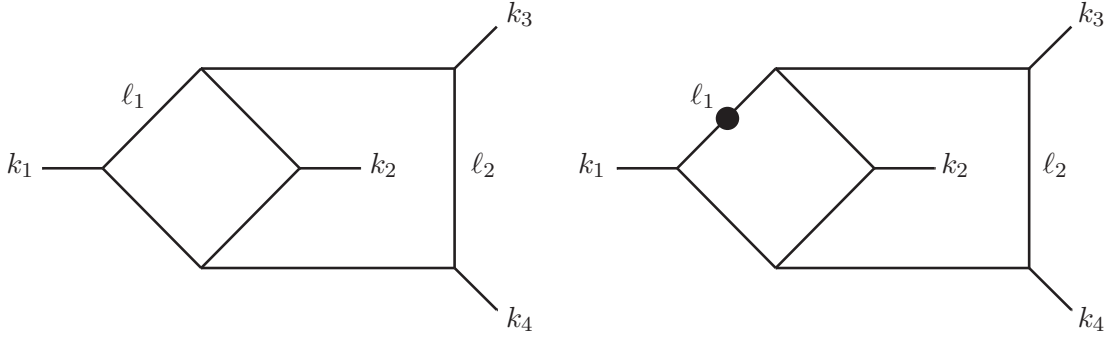

The nonplanar double box amplitude contribution with four massless external
legs has already been worked out in detail for an integral basis with a scalar
integral and a rank 1 tensor \cite{Badger:2012dp,Sogaard:2013yga}. However,
the reduction identities of the preceding subsection suggest an equivalent
integral basis in which the tensor integral is eliminated. We thus project the
amplitude onto two scalar integrals. Our master equation reads (see also
fig.~\ref{XBOX_SUBBOX})
\begin{align}
\label{XBOX_SCALAR_BASIS}
\mc A^{(2)}_4 = 
C_1\X(1,\dots,1,0,0)+
C_2\X(2,1,\dots,1,0,0)+\cdots
\end{align}
and we thus seek to determine $C_1$ and $C_2$.

\noindent
It is not hard to show that the master integral projectors in this basis are
\begin{align}
\mc M_1 = {} &
s_{12}^3
(-2(-1+2\chi^2),\,1+2\chi^2,\,-3+2\chi^2,\,-2(-1+2\chi^2),\, \nn \\
&\;\;\qquad\qquad 1+2\chi^2,\,-3+2\chi^2,\,1+2\chi^2,\,-3+2\chi^2,\,
1+2\chi^2,\,-3+2\chi^2)\;,\, \\[2mm]
\mc M_2 = {} &
\chi s_{12}^4
(-2(1+2\chi),\,-1+2\chi,\,3+2\chi,\,-2(1+2\chi),\, \nn \\
&\;\;\qquad\qquad -1+2\chi,\,3+2\chi,\,-1+2\chi,\,3+2\chi,\,-1+2\chi,\,3+2\chi)\;,
\end{align}
and the master integral coefficients in eq.~\eqref{XBOX_SCALAR_BASIS} can then be
written very compactly
\begin{align}
C_i = -\frac{1}{16s_{12}^3}\oint_{\mc M_i}\frac{dz}{z(z-\chi)(z-\chi-1)}
\sum_{\substack{\text{helicities}\\\text{particles}}}
\prod_{k=1}^6 A_{(k)}^\tree(z)\;.
\end{align}
The coefficients produced here agree in $D = 4$ with those worked out in
refs.~\cite{Badger:2012dp,Sogaard:2013yga} as can be verified using the IBP
identity \eqref{XBOXIBP1}.

\section{Discussion}
In this paper, we naturally generalized the maximal unitarity method to
integrals with doubled propagators and provided a simple way of reducing
integrals with doubled (or even higher-order) propagators onto a master
integral basis. The residues of an integral with doubled propagators are
degenerate, which cannot be directly calculated by Cauchy's theorem but can be
evaluated by computational algebraic geometry methods (Gr\"obner basis). Then
from the projector information, we obtain the master integral coefficients.
This method has been successfully tested on several one-loop and two-loop
examples.

Since the contour projector can be found by using IBPs without doubled
propagators, our method implies that the complete set of IBPs (involving
integrals with or without doubled propagators) can be derived from the set of
IBPs without doubled propagators. Our method can also be used for converting
between different integral bases.

So far, the maximal unitarity method for two-loop and higher-loop has been
tested only for diagrams in the $D=4$ limit. Therefore our paper only obtained
the finite part of the reduction of integrals with doubled propagators, but
not the $O(\epsilon)$ contribution.

On the other hand, in all our examples, the reduction coefficients of integral
with doubled propagators are finite, i.e., without poles in $\epsilon$. It is
not accidental: consider an integrand without doubled propagators 
$N/(f_1 \cdots f_k)$ in $n$ variables.  If (1) its cut solution is $n-k$
dimensional, (2) the cut can be parameterized by a set of variables $z$,
(3) the integration variables can be chosen to be $x$ and $z$, then
we evaluate the multivariate residue regarding the ideal 
$I(z)=\langle f_1(x;z),\dots,f_k(x;z)\rangle$ in  $x$, parameterized by $z$,
\begin{equation}
 \label{eq:15}
 \oint dz\ \oint d x\ \frac{\  N}{f_1 \ldots f_k} = \oint
 dz\ \frac{h(z)}{g(z)}\;.
\end{equation}
For integrals with doubled propagators, a similar calculation
regarding the ideal 
$\tilde I(z)=\langle f_1(x; z)^2,\dots,f_k(x;z)\rangle$ gives
\begin{equation}
  \label{eq:16}
  \oint dz\ \oint d x\ \frac{\  N}{f_1^2 \ldots f_k} = \oint
  dz\ \frac{\tilde h(z)}{\tilde g(z)}\;.
\end{equation}
Note that $g(z_0)=0$ if and only if the ideal $I(z_0)$ is not
zero-dimensional. Since the ideals $I(z_0)$ and $\tilde I(z_0)$ have
the same zero locus, $g(z_0)=0$ if and only if $\tilde g(z_0)=0$. Thus,
$\tilde g$ and $g$ have the same zeros in $z$, just different multiplicities.
Therefore the integral with doubled propagators does not generate new poles in
$z$, and the residues are still finite. Hence the reduction coefficients are
finite in the $D=4$ limit.

There are several promising future directions. We expect that the maximal
unitarity method (including integrals with doubled propagators) can be
generalized to $D$-dimensional cases by a contour integral in the extra
dimension and analytic continuation in $D$. Moreover, the reduction algorithm
for integrals with higher powers of propagators should apply seamlessly to
massive external legs.

For the computational aspect, the multivariate residue calculation can be sped
up by using the relation between multivariate residues and the Bezoutian matrix
\cite{MR2161985,Sogaard:2014oka}. Then we do not need to find the Gr\"obner
basis in the lexicographic order.

\acknowledgments
We are grateful to Emil Bjerrum-Bohr, Simon Caron-Huot, Poul Henrik Damgaard,
Tristan Dennen, Rijun Huang and David Kosower for useful discussions. It is a
pleasure to thank Simon Badger and Hjalte Frellesvig for comments on the
manuscript in draft stage. Both authors acknowledge the organizers and
participants of the workshop {\it The Geometry and Physics of Scattering
Amplitudes} at Simons Center for Geometry and Physics, Stony Brook University.
We also express gratitude to Institut de Physique Th\'eorique, CEA Saclay, and
in particular David Kosower for hospitality during this project. The work of
YZ is supported by Danish Council for Independent Research (FNU) grant
11-107241.

\clearpage
\appendix
\section{Planar Double Box Hepta-Cuts}
\label{DBOXDOUBLEDCUTS}
For the sake of completeness, we include all hepta-cuts of four-point planar
double box scalar integrals with a doubled propagator. The ordering of the
propagators follows that of the main text.
\begin{align}
\P(2,1,\dots,1,0,0)_{\mc S_{1,3}} = {} & 
-\frac{1}{16s_{12}^4}\oint\frac{dz}{z(z+\chi)} \\[2mm]
\P(2,1,\dots,1,0,0)_{\mc S_{2,4,5,6}} = {} & 
-\frac{\chi}{16s_{12}^4}\oint\frac{dz}{z(z+\chi)^2}
\\[5mm]
\P(1,2,1,\dots,1,0,0)_{\mc S_{1,3}} = {} & 
-\frac{1}{16s_{12}^4}\oint\frac{dz}{z(z+\chi)^2} \\[2mm]
\P(1,2,1,\dots,1,0,0)_{\mc S_{2,4,5,6}} = {} & 
+\frac{1}{16s_{12}^4}\oint dz\frac{\chi(1+\chi)+z(1+2\chi)}
{z^2(z+\chi)^2}
\\[5mm]
\P(1,1,2,1,\dots,1,0,0)_{\mc S_{1,3}} = {} & 
-\frac{1}{16s_{12}^4}\oint\frac{dz}{z(z+\chi)} \\[2mm]
\P(1,1,2,1,\dots,1,0,0)_{\mc S_{2,4,5,6}} = {} & 
-\frac{\chi}{16s_{12}^4}\oint\frac{dz}{z(z+\chi)^2}
\\[5mm]
\P(1,1,1,2,1,1,1,0,0)_{\mc S_{1,3}} = {} & 
-\frac{\chi}{16s_{12}^4}\oint\frac{dz}{z(z+\chi)^2} \\[2mm]
\P(1,1,1,2,1,1,1,0,0)_{\mc S_{2,4,5,6}} = {} & 
-\frac{1}{16s_{12}^4}\oint\frac{dz}{z(z+\chi)}
\\[5mm]
\P(1,\dots,1,2,1,1,0,0)_{\mc S_{1,3}} = {} & 
+\frac{1}{16s_{12}^4}\oint dz\frac{\chi(1+\chi)+z(1+2\chi)}
{z^2(z+\chi)^2} \\[2mm]
\P(1,\dots,1,2,1,1,0,0)_{\mc S_{2,4,5,6}} = {} & 
-\frac{1}{16s_{12}^4}\oint\frac{dz}{z(z+\chi)^2}
\\[5mm]
\P(1,\dots,1,2,1,0,0)_{\mc S_{1,3}} = {} & 
-\frac{\chi}{16s_{12}^4}\oint\frac{dz}{z(z+\chi)^2}\;, \\[2mm]
\P(1,\dots,1,2,1,0,0)_{\mc S_{2,4,5,6}} = {} & 
-\frac{1}{16s_{12}^4}\oint\frac{dz}{z(z+\chi)}
\\[5mm]
\P(1,\dots,1,2,0,0)_{\mc S_i} = {} & 
-\frac{1}{16s_{12}^4}\oint\frac{dz}{z(z+\chi)^2}
\end{align}

\clearpage
\section{Nonplanar Double Box Hepta-Cuts}
\label{XBOXDOUBLEDCUTS}
We also provide explicit forms of the hepta-cuts of all four-point nonplanar
double box integrals with a single doubled propagator and a scalar numerator.
The overall signs are determined by consistency of relations among residues.
\begin{align}
\X(2,1,\dots,1,0,0)_{\mc S_{3,4}} = {} & 
+\frac{1}{16s_{12}^4}\oint dz\frac{\chi+(1+\chi)z}{z^2(z+\chi)^2} \\[2mm]
\X(2,1,\dots,1,0,0)_{\mc S'_{7,8}} = {} &
+\frac{1}{16s_{12}^4}\oint dz\frac{1+\chi}{z^2(z-\chi-1)} \\[2mm]
\X(2,1,\dots,1,0,0)_{\mc S'_{1,2,5,6}} = {} &
+\frac{1}{16s_{12}^4}\oint dz\frac{1}{z(z-\chi)^2(z-\chi-1)}
\\[5mm]
\X(1,2,1\dots,1,0,0)_{\mc S'_{3,4}} = {} & 
-\frac{1}{16s_{12}^4}\oint dz\frac{\chi}{z^2(z+\chi)}\;, \\[2mm]
\X(1,2,1,\dots,1,0,0)_{\mc S'_{7,8}} = {} &
-\frac{1}{16s_{12}^4}\oint dz\frac{1+\chi(1+z)}{z^2(z-\chi-1)^2} \\[2mm]
\X(1,2,1,\dots,1,0,0)_{\mc S'_{1,2,5,6}} = {} &
-\frac{1}{16s_{12}^4}\oint dz\frac{1}{z(z-\chi)(z-\chi-1)^2}
\\[5mm]
\X(1,1,2,1,\dots,1,0,0)_{\mc S'_{3,4}} = {} & 
-\frac{1}{16s_{12}^4}\oint dz\frac{2\chi+z}{z(z+\chi)^2} \\[2mm]
\X(1,1,2,1,\dots,1,0,0)_{\mc S'_{7,8}} = {} &
+\frac{1}{16s_{12}^4}\oint dz\frac{2(1+\chi)-z}{z(z-\chi-1)^2} \\[2mm]
\X(1,1,2,1,\dots,1,0,0)_{\mc S'_{1,2,5,6}} = {} &
-\frac{1}{16s_{12}^4}\oint dz\frac{2\chi(1+\chi-z)-z}{z(z-\chi)^2(z-\chi-1)^2}
\\[5mm]
\X(1,1,1,2,1,1,1,0,0)_{\mc S'_{3,4}} = {} & 
-\frac{1}{16s_{12}^4}\oint dz\frac{1}{z(z+\chi)^2} \\[2mm]
\X(1,1,1,2,1,1,1,0,0)_{\mc S'_{7,8}} = {} &
-\frac{1}{16s_{12}^4}\oint dz\frac{1}{z(z-\chi-1)^2} \\[2mm]
\X(1,1,1,2,1,1,1,0,0)_{\mc S'_{1,2,5,6}} = {} &
-\frac{1}{16s_{12}^4}\oint dz\frac{
(1+2(\chi-z))(\chi(1+\chi)-(1+2\chi)z)
}{z^2(z-\chi)^2(z-\chi-1)^2} \\ 
\X(1,\dots,1,2,1,1,0,0)_{\mc S'_{3,4}} = {} & 
-\frac{1}{16s_{12}^4}\oint dz\frac{2\chi+z}{z(z+\chi)^2} \\[2mm]
\X(1,\dots,1,2,1,1,0,0)_{\mc S'_{7,8}} = {} &
+\frac{1}{16s_{12}^4}\oint dz\frac{2(1+\chi)-z}{z(z-\chi-1)^2} \\[2mm]
\X(1,\dots,1,2,1,1,0,0)_{\mc S'_{1,2,5,6}} = {} &
+\frac{1}{16s_{12}^4}\oint dz\frac{2\chi(z-\chi-1)+z}{z(z-\chi)^2(z-\chi-1)^2}
\end{align}
\begin{align}
\X(1,\dots,1,2,1,0,0)_{\mc S'_{3,4}} = {} & 
-\frac{1}{16s_{12}^4}\oint dz\frac{2\chi+z}{z(z+\chi)^2} \\[2mm]
\X(1,\dots,1,2,1,0,0)_{\mc S'_{7,8}} = {} &
+\frac{1}{16s_{12}^4}\oint dz\frac{2(1+\chi)-z}{z(z-\chi-1)^2} \\[2mm]
\X(1,\dots,1,2,1,0,0)_{\mc S'_{1,2,5,6}} = {} &
+\frac{1}{16s_{12}^4}\oint dz\frac{2\chi(z-\chi-1)+z}{z(z-\chi)^2(z-\chi-1)^2}
\\[5mm]
\X(1,\dots,1,2,0,0)_{\mc S'_{3,4}} = {} & 
-\frac{1}{16s_{12}^4}\oint dz\frac{\chi(1-\chi)+(1+\chi)z}{z(z+\chi)^3} \\[2mm]
\X(1,\dots,1,2,0,0)_{\mc S'_{7,8}} = {} &
-\frac{1}{16s_{12}^4}\oint dz\frac{1+\chi}{z(z-\chi-1)^2}\;, \\[2mm]
\X(1,\dots,1,2,0,0)_{\mc S'_{1,2,5,6}} = {} &
+\frac{1}{16s_{12}^4}\oint dz\frac{1}{z(z-\chi)^2(z-\chi-1)}
\end{align}

\end{document}